\documentclass[pra,twocolumn,shownopacs,amssymb]{revtex4}
\usepackage{graphicx}
\usepackage{amsmath}

\begin{document}

\title{Superfluid stiffness for the attractive Hubbard model on a honeycomb optical lattice}

\author{M. Iskin}
\affiliation{Department of Physics, Ko\c{c} University, Rumelifeneri Yolu, 
34450 Sar\i yer, Istanbul, Turkey}

\date{\today}

\begin{abstract}

In addition to the conventional contribution that is directly controlled by the single-particle 
energy spectrum, the superfluid phase stiffness of a two-component Fermi gas has 
a geometric contribution that is governed by the quantum metric of the honeycomb's 
band structure. Here, we take both contributions into account, and construct phase 
diagrams for the critical superfluid transition temperature as a function of the chemical 
potential, particle filling, onsite interaction and next-nearest-neighbor hopping. 
Our theoretical approach is based on a self-consistent solution of the BCS mean-field 
theory for the stationary Cooper pairs and the universal BKT relation for the phase 
fluctuations.

\end{abstract}


\maketitle

\section{Introduction}
\label{sec:intro}

Following the pioneering works by Peotta and T\"orm\"a on the origins of 
superfluidity in topologically nontrivial flat bands~\cite{torma15}, the 
deeper connection between some of the superfluid (SF) properties of a 
two-component Fermi gas and the quantum geometry of its non-interacting 
Bloch bands came as a complete surprise in recent 
years~\cite{torma16, torma17a, iskin17, iskin18a, iskin18b}. It has been 
found on general grounds that the SF weight of a multi-band SF with a 
uniform order parameter can be separated into two distinct parts, depending 
on the physical mechanisms involved. While the real intraband processes are 
attributed to the conventional contribution, the virtual interband processes 
are attributed to the geometric one. Alternatively, in contrast to the conventional 
contribution that is solely controlled by the derivatives of the energy dispersions, 
the geometric one is also associated with the derivatives of the underlying 
Bloch wave functions~\cite{torma17a}. For instance, unless the geometric 
interband contribution vanishes, superfluidity prevails in a flat band thanks 
to the presence of other flat or dispersive bands~\cite{torma15, torma16}. 
More recently, the root cause of this deeper connection has been identified as a 
mass-renormalization mechanism for the SF carriers, i.e., the quantum 
geometry governs not only the SF weight but also some other SF properties 
through renormalizing the effective mass of the two-body bound states 
and of Cooper pairs in general~\cite{iskin17, iskin18b}.

Furthermore, in the particular cases of flat-band and two-band systems, the 
geometric contribution to the SF weight is simply controlled by the so-called quantum 
metric~\cite{torma15, torma16, torma17a, iskin17, iskin18a}. The quantum metric 
corresponds to the real part of the quantum geometric tensor, and its 
geometrical importance reveals itself as a measure of the quantum distance 
between nearby Bloch states~\cite{provost80, berry89, thouless98}. 
Note that the imaginary part of the quantum geometric tensor corresponds 
to the so-called Berry curvature, which is a distinct but related quantity 
associated with the emergent gauge field in momentum space, i.e., 
characterizing its quantum topology~\cite{provost80, berry89, thouless98}. 
Some of the two-band SFs that have already been analyzed in this 
context are the Haldane-Hubbard model~\cite{torma17a}, Kane-Mele-Hubbard
model~\cite{torma17a}, time-reversal-invariant Hofstadter-Hubbard 
model~\cite{torma15, iskin17}, and the spin-orbit coupled Fermi 
gases~\cite{iskin18a}. These works show clear signs that understanding 
the quantum metric effects on any one of these models may eventually have 
far reaching implications for a wide class of two-band SFs.

Motivated by these theoretical proposals as well as ongoing experimental 
efforts utilizing cold Fermions on various forms of honeycomb optical 
structures~\cite{tarruell12, uehlinger12, polini13, jotzu14, flaschner16}, 
here we calculate the critical SF transition temperature of the attractive Hubbard 
model on a two-dimensional honeycomb lattice for a large window of 
model parameters. Our theoretical approach is based on a self-consistent 
solution of the BCS mean-field theory for the stationary Cooper pairs and 
the universal BKT relation for the phase fluctuations, and we have two main 
goals. In addition to constructing the phase diagrams for the critical SF transition
temperature, we plan to uncover the critical role played by the quantum 
geometry of the underlying band structure. 
For instance, while the highest attainable critical temperature is found to be 
around $0.15t$ for the nearest-neighbor-hopping model, it increases quite 
rapidly with the inclusion of next-nearest-neighbor hoppings. In addition, the
relative weight of the quantum metric contribution to the SF phase stiffness 
is found to be a non-monotonous function of the interaction strength, and 
it may reach beyond $\%50$ depending on the parameters. 
Thus, these findings arguably suggest that a SF Fermi gas that is loaded 
on a honeycomb lattice is one of the ideal platforms for studying 
quantum geometric effects with cold atoms. 

The rest of the paper is organized as follows. The theoretical framework is 
presented in Sec.~\ref{sec:ta}, where we first discuss the honeycomb's band 
structure and highlight the presence of Dirac cones in Sec.~\ref{sec:bs}, 
then introduce the BCS mean-field theory and derive the order-parameter 
and number equations in Sec.~\ref{sec:bcs}, 
and then review the BKT relation and the SF stiffness 
in Sec.~\ref{sec:bkt}. Having a complete set of self-consistency equations for 
determining the critical SF transition temperature, we present its numerical 
analysis in Secs.~\ref{sec:nr} and~\ref{sec:disc}, and conclude the paper with 
our final remarks in Sec.~\ref{sec:conc}.

\section{Theoretical Approach}
\label{sec:ta}

\begin{figure}[htbp]
\includegraphics[scale=0.25]{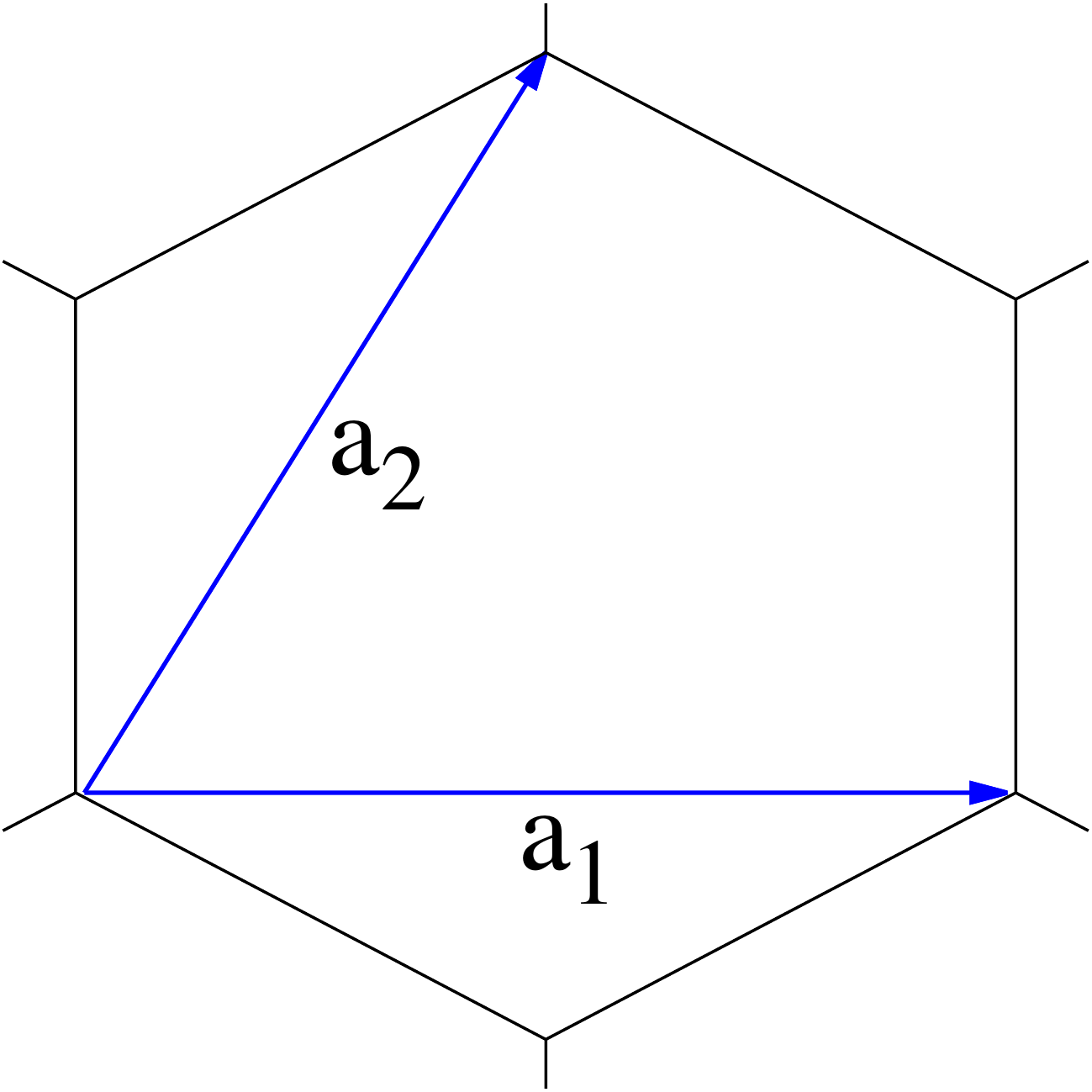}
\caption{\label{fig:honeycomb}
Primitive lattice vectors for the honeycomb lattice.
}
\end{figure}

The honeycomb lattice is a two-dimensional crystal structure with a hexagonal 
Bravais lattice and a two-site basis. In this paper, we denote its lattice spacing 
by $a$, and choose
$
\mathbf{a_1} = (\sqrt{3}a, 0)
$
and
$
\mathbf{a_2} = (\sqrt{3}a/2, 3a/2)
$
as the primitive lattice vectors for its Bravais lattice as shown in Fig.~\ref{fig:honeycomb}. 
The corresponding reciprocal lattice vectors
$
\mathbf{b_1} = [2\pi/(\sqrt{3}a), -2\pi/(3a)] 
$
and
$
\mathbf{b_2} = [0, 4\pi/(3a)]
$
also form a hexagonal lattice in reciprocal space, leading to a first Brillouin zone 
that has the shape of a hexagon with side-length $4\pi/(3\sqrt{3}a)$.
Due to its two-site basis on a Bravais lattice, a honeycomb lattice gives rise to 
a two-band structure with important features for the single-particle problem. 
For the sake of completeness, let us first discuss its band structure and highlight 
the presence of Dirac cones, as they turn out to play critical roles in the 
many-body problem as well.

\subsection{Band Structure}
\label{sec:bs}

Within the tight-binding approximation, the single-particle Hamiltonian can be 
written as
$
H_\sigma = - \sum_{i \in S, j \in S'} t_{S i, S' j} c^\dagger_{\sigma S i} c_{\sigma S' j},
$
where the pseudo-spin $\sigma \equiv \{\uparrow, \downarrow\}$ denotes the two
components of a Fermi gas, the index $i \in S$ refers to a site $i$ in the hexagonal 
sublattice $S \equiv \{A, B\}$, the hopping parameter $t_{S i, S' j}$ characterizes the 
tunneling amplitude from site $j$ to $i$, and the operator $c^\dagger_{\sigma S i}$ 
($c_{\sigma S i}$) creates (annihilates) a $\sigma$ particle on $i \in S$. 
In this paper, we set the nearest-neighbor (i.e.,  inter-sublattice) hopping parameter
$t$ as our energy scale, and vary the next-nearest-neighbor (i.e.,  intra-sublattice) 
one $t'$ accordingly.

Using the Fourier expansion of the creation and annihilation operators in the 
reciprocal ($\mathbf{k}$) space, e.g.,
$
c_{\sigma S i} = (1/\sqrt{M_S}) \sum_\mathbf{k} c_{\sigma S \mathbf{k}} 
e^{\textrm{i} \mathbf{k} \cdot \mathbf{r_i}}
$
where $M_A = M_B = M/2$ is the number of sites in the hexagonal sublattice,
a compact way to express this Hamiltonian is
$
H_\sigma = \sum_\mathbf{k} \psi_{\sigma \mathbf{k}}^\dagger h_\mathbf{k} \psi_{\sigma \mathbf{k}}.
$
Here, the creation operator
$
\psi_{\sigma \mathbf{k}}^\dagger = (c_{\sigma A \mathbf{k}}^\dagger \, c_{\sigma B \mathbf{k}}^\dagger)
$
is in the form of a two-component sublattice spinor with the Hamiltonian density
$
h_\mathbf{k} = d_\mathbf{k}^0 \tau_0 + \mathbf{d_\mathbf{k}} \cdot \boldsymbol{\tau},
$
where $\tau_0$ is a unit matrix and $\boldsymbol{\tau} = (\tau_x, \tau_y)$ is a vector 
of Pauli matrices. Note that while the diagonal element
$
d_\mathbf{k}^0 = -2t' \cos(\sqrt{3} k_x a) - 4t' \cos(\sqrt{3} k_x a/2) \cos(3 k_y a/2)
$
of $h_\mathbf{k}$ is due solely to the next-nearest-neighbor hoppings, the off-diagonal
element $\mathbf{d_k} = (d_\mathbf{k}^x, d_\mathbf{k}^y)$ with
$
d_\mathbf{k}^x = -t \cos(k_y a) - 2t \cos(k_y a/2) \cos(\sqrt{3} k_x a/2)
$
and
$
d_\mathbf{k}^y = t \sin(k_y a) - 2t \sin(k_y a/2) \cos(\sqrt{3} k_x a/2)
$
is due solely to the nearest-neighbor ones. 

Thus, the single-particle energy eigenvalues are simply determined by 
$
\varepsilon_{s \mathbf{k}} = d_\mathbf{k}^0 + s d_\mathbf{k},
$
where $s = \pm$ denotes the upper/lower band and $d_\mathbf{k} = |\mathbf{d_k}|$ 
reduces to $t \sqrt{3 - d_\mathbf{k}^0/t'}$. It can be shown that both energy bands exhibit 
a total of two Dirac cones that are equally distributed among the six corners of 
the first Brillouin zone. Since $d_\mathbf{k}^0 \to 3t'$ and $d_\mathbf{k} \to 0$ at the tips 
of the cones, the density $\mathcal{N}_\varepsilon$ of single-particle states 
vanishes at $\varepsilon = 3t'$, i.e., where the upper and lower cones touch 
each other. This analysis suggests that the low-temperature behavior of a 
Fermi gas that is loaded on a two-dimensional honeycomb lattice is directly 
controlled by the presence of these Dirac cones in the band structure. 
For instance, the Fermi gas shows a semi-metallic behavior that persists up to 
a critical interaction threshold depending on the temperature~\cite{zhao06, lee09}. 

Our main goals in this work are twofold. We would like not only to construct the 
phase diagrams for the critical SF transition temperature of the attractive Hubbard 
model on a honeycomb lattice, but also to uncover the critical role played by the 
quantum geometry of the underlying band structure. These are achieved by 
adapting a self-consistent approach that is based on the simultaneous solution
of the BCS mean-field theory for the stationary Cooper pairs and the universal 
BKT relation for the phase fluctuations.

\subsection{BCS Mean-field Theory}
\label{sec:bcs}

For a two-component Fermi gas that is considered in this work, the attractive Hubbard 
model can be written as $H = \sum_\sigma H_\sigma + H_\textrm{int} + H_\mu$, where
$
H_\textrm{int} = - U \sum_{S i} c_{\uparrow S i}^\dagger c_{\downarrow S i}^\dagger 
c_{\downarrow S i} c_{\uparrow S i}
$
with $U \ge 0$ takes the onsite intercomponent interactions into account. We treat the 
interaction term within the BCS mean-field approximation for pairing, and characterize 
various SF phases through the complex order parameter
$
\Delta_{S i} = U \langle c_{\downarrow S i} c_{\uparrow S i} \rangle,
$
where $\langle \ldots \rangle$ denotes the thermal average. However, thanks to the 
time-reversal symmetry of $H$, $\Delta_{S i}$ turns out to be uniform for a given 
sublattice, i.e.,
$
\Delta_S = (1/M_S) \sum_{i \in S} \Delta_{S i}.
$
Furthermore, in order to fix the total number 
$
N = \sum_{\sigma S i} \langle c_{\sigma S i}^\dagger c_{\sigma S i} \rangle
$
of particles in the thermal state, we include an additional term
$
H_\mu = - \mu \sum_{\sigma S i} c_{\sigma S i}^\dagger c_{\sigma S i}
$
in $H$, where $\mu$ is the chemical potential.

Similar to the single-particle problem, a compact way to express the mean-field 
Hamiltonian is 
\begin{align}
H_\textrm{mf} = C + \sum_\mathbf{k} \Psi_\mathbf{k}^\dagger 
\left(
\begin{array}{cc}
  h_\mathbf{k} - \mu \tau_0& \boldsymbol{\Delta} \\
  \boldsymbol{\Delta}^\dagger  & -h_{-\mathbf{k}}^* + \mu \tau_0 \\
\end{array}
\right)
\Psi_\mathbf{k},
\label{eqn:ham}
\end{align}
where
$
C =  \sum_\mathbf{k} \mathrm{Tr} \lbrace h_{-\mathbf{k}} - \mu \tau_0 
+ \boldsymbol{\Delta}^\dagger \boldsymbol{\Delta} /U \rbrace
$
is a constant with $\mathrm{Tr}$ denoting the trace over sublattices, 
$
\Psi_\mathbf{k}^\dagger = (\psi_{\uparrow \mathbf{k}}^\dagger \, \psi_{\downarrow, -\mathbf{k}}^\dagger)
$
is a four-component spinor operator, and 
$
\boldsymbol{\Delta}_{S S'} = \Delta_S \delta_{S S'}
$
with $\delta_{ij}$ the Kronecker-delta is diagonal in the sublattice sector. 
Since 
$
\Delta_S = (U/M_S) \sum_{\mathbf{k}} 
\langle c_{\downarrow S,-\mathbf{k}} c_{\uparrow S \mathbf{k}} \rangle
$
turns out to be uniform for the entire lattice thanks to the inversion symmetry of 
the $A$ and $B$ sublattices, we take $\Delta_A = \Delta_B = \Delta$ 
as real without losing generality. Combining this expression with the 
number equation
$
N = \sum_{\sigma S \mathbf{k}}
\langle c_{\sigma S \mathbf{k}}^\dag c_{\sigma S \mathbf{k}} \rangle,
$
we eventually obtain a set of self-consistency equations
\begin{align}
\label{eqn:op}
1 &= \frac{U}{2M} \sum_{s \mathbf{k}} \frac{\mathcal{X}_{s \mathbf{k}}}{E_{s \mathbf{k}}}, \\
F &= 1 - \frac{1}{M} \sum_{s \mathbf{k}} 
\frac{\mathcal{X}_{s \mathbf{k}}}{E_{s\mathbf{k}}} \xi_{s \mathbf{k}},
\label{eqn:filling}
\end{align}
for $\Delta$ and $\mu$. Here, $M$ is the number of lattice sites,
$
\xi_{s \mathbf{k}}=\varepsilon_{s \mathbf{k}} - \mu
$
is the shifted dispersion,
$
\mathcal{X}_{s \mathbf{k}} = \tanh [E_{s \mathbf{k}}/(2T)]
$
is a thermal factor with $k_\textrm{B} \to 1$ the Boltzmann constant and $T$ the 
temperature,
$
E_{s \mathbf{k}} = \sqrt{\xi_{s \mathbf{k}}^2 + \Delta^2}
$
is the quasi-particle energy spectrum, and $0 \le F = N/M \le 2$ is the total particle filling. 
Thus, we use Eqs.~(\ref{eqn:op}) and~(\ref{eqn:filling}) to determine $\Delta$ and 
$\mu$ for any given set of $U$, $F$, $T$ and $t'$ parameters. 

While the critical BCS transition temperature $T_\textrm{BCS}$ is simply determined 
by setting $\Delta \to 0$ in Eqs.~(\ref{eqn:op}) and (\ref{eqn:filling}), the mean-field 
theory is known to give qualitatively reliable results for $U \lesssim t$ only. 
This is because growing phase fluctuations eventually break the mean-field 
approximation down in the $U \gg t$ limit, for which $T_\textrm{BCS} \propto U$ 
characterizes the pair formation temperature~\cite{nsr85, randeria92}. 
In particular to two dimensions, the SF phase coherence temperature is 
determined by the universal BKT relation, leading to a much lower result.

\subsection{BKT Temperature and Superfluid Stiffness}
\label{sec:bkt}

Going beyond the mean-field theory, and including the phase fluctuations, the critical 
SF transition temperature is determined by the universal BKT relation through an 
analogy with the XY model~\cite{b, kt, nk}. This approach has long been applied 
to the single-band SFs with great success~\cite{denteneer93}, and it has recently 
been generalized to the case of multi-band SFs with uniform order 
parameters~\cite{torma16, torma17a, iskin17}. In the case of two-band SFs, one 
finds~\cite{torma17a, iskin18a}
\begin{align}
\label{eqn:bkt}
T_\textrm{BKT} &= \frac{\pi}{8} \sqrt{\det{\boldsymbol{D}}},\\
\label{eqn:conv}
\boldsymbol{D}_{\mu\nu}^\textrm{conv} & = \frac{\Delta^2}{\mathcal{A}} \sum_{s\mathbf{k}} 
\left(
\frac{\mathcal{X}_{s\mathbf{k}}}{E_{s\mathbf{k}}^3} 
- \frac{\mathcal{Y}_{s\mathbf{k}}}{2T E_{s\mathbf{k}}^2}
\right)
\frac{\partial \xi_{s\mathbf{k}}} {\partial k_\mu}
\frac{\partial \xi_{s\mathbf{k}}} {\partial k_\nu}, \\
\boldsymbol{D}_{\mu\nu}^\textrm{geom} & = - \frac{2\Delta^2}{\mathcal{A}} \sum_{s \mathbf{k}}
\frac{d_\mathbf{k} \mathcal{X}_{s\mathbf{k}}}{s (d_\mathbf{k}^0 - \mu) E_{s\mathbf{k}}} 
g_{\mu\nu}^\mathbf{k},
\label{eqn:geom}
\end{align}
where 
$
\boldsymbol{D}_{\mu\nu} = \boldsymbol{D}_{\mu\nu}^\textrm{conv} 
+ \boldsymbol{D}_{\mu\nu}^\textrm{geom}
$ 
is the SF phase stiffness, $\mathcal{A}$ is the area of the system, and 
$
\mathcal{Y}_{s\mathbf{k}} = \mathrm{sech}^2 [E_{s\mathbf{k}}/(2T)]
$
is a thermal factor. Here, while the conventional contribution 
$\boldsymbol{D}_{\mu\nu}^\textrm{conv}$ is of the usual single-band form 
accounting for real intraband processes, the geometric contribution 
$\boldsymbol{D}_{\mu\nu}^\textrm{geom}$ is due to virtual interband processes 
that are directly controlled by the total quantum metric of the single-particle 
bands, i.e., 
$
g_{\mu\nu}^\mathbf{k} = 
(\partial \widehat{\mathbf{d}}_\mathbf{k} / \partial k_\mu)  \cdot 
(\partial \widehat{\mathbf{d}}_\mathbf{k} / \partial k_\nu) / 2
$
with 
$
\widehat{\mathbf{d}}_\mathbf{k} = \mathbf{d}_\mathbf{k}/d_\mathbf{k}
$
a unit vector. 

In contrast to the standard BCS mean-field theory, it turns out that a self-consistent 
solution of Eqs.~(\ref{eqn:op})-(\ref{eqn:geom}) for $\Delta(T_\textrm{BKT})$, 
$\mu(T_\textrm{BKT})$ and $T_\textrm{BKT}$ provides a qualitatively reliable 
description of the critical SF transition temperature for both $U \lesssim t$ and 
$U \gg t$ limits~\cite{denteneer93}. 
Even though this approach is still far from being quantitatively accurate
in comparison to the numerically-exact ones~\cite{lee09}, its much simpler 
analytical construction provides considerable insight into the main features.
For instance, Eq.~(\ref{eqn:bkt}) puts $T_\textrm{BCS}$ as the upper bound 
on $T_\textrm{BKT}$ in such a way that $T_\textrm{BKT} \to T_\textrm{BCS}$ 
from below when $U \lesssim t$, and that $T_\textrm{BKT}  \ll T_\textrm{BCS}$ 
when $U \gg t$. The latter limit can be shown by noting that
$
\Delta = (U/2) \sqrt{F(2-F)},
$
$
\mu = -(U/2) (1-F),
$
and
$
\boldsymbol{D}_{\mu\nu} = \lbrace \Delta^2/[\mathcal{A} (\mu^2+\Delta^2)^{3/2}]\rbrace 
\sum_\mathbf{k} \textrm{Tr} \lbrace
(\partial h_\mathbf{k} / \partial k_\mu)
(\partial h_\mathbf{k} / \partial k_\nu) 
\rbrace,
$
leading to a diagonal SF stiffness $\boldsymbol{D}_{\mu\nu} = D_0 \delta_{\mu\nu}$ with
$
D_0 = 2\Delta^2 (t^2 + 6t'^2) /[\sqrt{3} (\mu^2+\Delta^2)^{3/2}].
$
Thus, the BKT relation~(\ref{eqn:bkt}) gives
$
T_\textrm{BKT} = \pi F (2-F) (t^2 + 6t'^2)/(2\sqrt{3} U),
$
showing that, independently of its sign, $t'$ increases $T_\textrm{BKT}$ 
for a given $F$ in the $U \gg t$ limit.
Except for the $U \lesssim t$ and $U \gg t$ limits, the self-consistency equations 
are not analytically tractable in general, and we resort to numerical methods 
instead.

\section{Numerical Results}
\label{sec:nr}

Having introduced the theoretical framework, here we implement an iterative 
numerical approach to find fully self-consistent solutions for 
$\Delta(T_\textrm{BKT})$, $\mu(T_\textrm{BKT})$ and
$\boldsymbol{D}(T_\textrm{BKT})$ that satisfy all 
Eqs.~(\ref{eqn:op})-(\ref{eqn:geom}) simultaneously for a given
set of model parameters. This allows us not only to construct phase diagrams 
for the critical SF transition temperature, but also to uncover the critical role 
played by the quantum geometry of the underlying band structure. For this 
purpose, next we choose a set of exemplary $t'/t \le 0$ ratios, and present the 
self-consistent results for $T_{BKT}/t$ and $D_0^\textrm{geom}/D_0$ as a
function of $\mu/t$, $F$ and $U/t$. Note that thanks to the apparent symmetry 
between the parameter sets $(t'/t > 0, \mu/t, F)$ and $(t'/t < 0, 6-\mu/t, 2-F)$, 
our phase diagrams cover the entire parameter regime of the model Hamiltonian.

\begin{figure*}[htbp]
\includegraphics[scale=0.7]{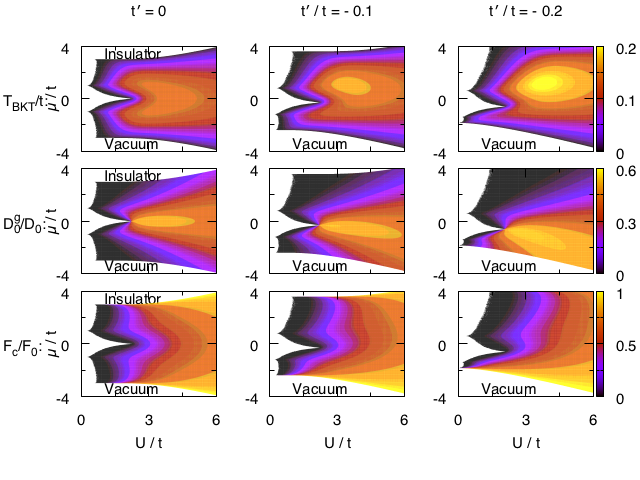}
\caption{(color online)
\label{fig:mu}
The critical SF transition temperature $T_\textrm{BKT}/t$ is shown in the 
upper row, the relative weight $D_0^\textrm{geom}/D_0$ of the geometric 
SF stiffness is shown in the middle row, and the faction $F_c/F_0$ of condensed 
particles is shown in the lower row.
}
\end{figure*}

\begin{figure*}[htbp]
\includegraphics[scale=0.7]{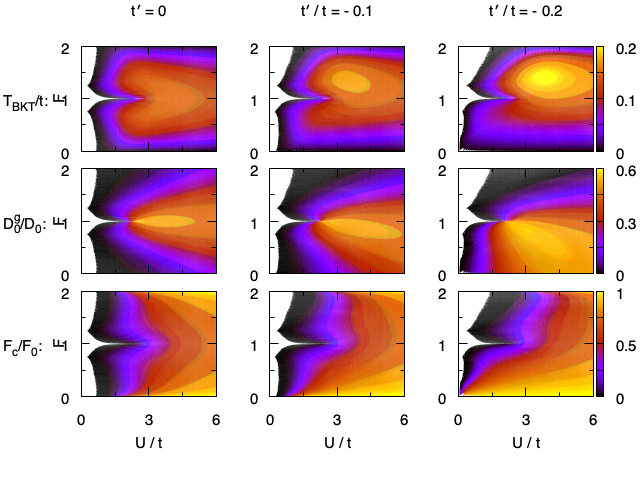}
\caption{(color online)
\label{fig:F}
Same as Fig.~\ref{fig:mu} but in the plane of total particle filling $F$.
}
\end{figure*}

Let us first set $U = 0$ and analyze the non-interacting limit. Since the lower 
single-particle band lies within the energy interval $-3t - 6t' \le \varepsilon \le 3t'$ 
and the upper one lies within $3t' \le \varepsilon \le 3t - 6t'$, the Fermi gas first 
fills the lower band as a function of increasing its Fermi energy $\varepsilon_F = \mu$ 
up until $\mu = 3t'$. For this reason while $\mu < -3t - 6t'$ region is denoted 
as a vacuum of particles with $F = 0$ and $\mu > 3t - 6t'$ region is denoted 
as a vacuum of holes (i.e., a band insulator) with $F = 2$, $\mu = 3t'$ 
corresponds precisely to the half filling with $F = 1$. Thus, except for the 
symmetric point $\mu = 3t'$ where the single-particle density of states 
$\mathcal{N}_\mu$ vanishes like a semi-metal, the ground state of a 
non-interacting Fermi gas is normal when $-3t - 6t' \le \mu \le 3t - 6t'$. 
All of these regions are clearly seen in Figs.~\ref{fig:mu} and~\ref{fig:F}.

Once the interactions are turned on, the BCS theory suggests that the normal 
region immediately transits into a SF with 
$
T_\textrm{BCS} \propto t e^{-1/(U \mathcal{N}_\mu)},
$
coinciding with $T_\textrm{BKT}$ in the $U \lesssim t$ limit. 
While such an exponential growth is clearly seen in the darker region in 
Figs.~\ref{fig:mu} and~\ref{fig:F}, our convergence scheme eventually fails in 
the white region when $U/t \to 0$, where $T_\textrm{BKT}$ becomes comparable 
to our relative numerical accuracy (i.e., $10^{-6}t$) between two consecutive 
iterations. In Fig.~\ref{fig:mu}, we note that the periphery of the white region 
nicely follows the general structure of $\mathcal{N}_\mu$, including 
its van Hove singularities at $F = 0.75$ and $1.25$. 
On the other hand, both the particle and hole vacuums as well as the semi-metal 
phase transit into a SF at finite interaction thresholds, beyond which 
$T_\textrm{BKT}/t$ grows as $\sqrt{U/U_c - 1}$ nearby the former regions and 
as $(U/U_c - 1)$ nearby $\mu = 3t'$. For instance, we find that the critical 
interaction threshold $U_c/t \approx \{ 2.23, 2.19, 2.06, 1.71 \}$ 
decreases with $t'/t = \{ 0, -0.1, -0.2, -0.3 \}$ 
for the semi-metal to SF phase transition at half filling. These are consistent with 
the known results in the literature where $U_c/t \approx \{2.24, 2.13\}$ for 
$t'/t = \{0, -0.15\}$~\cite{zhao06, cichy18}.
We emphasize that, in contrast to the normal to SF transition boundary, our vacuum, 
insulator and semi-metal to SF transition boundaries are very accurate as 
$T_\textrm{BKT}/t$ vanish quite rapidly near $U_c$.

In addition, for $t' = 0$, Fig.~\ref{fig:F} shows that a maximal value of 
$T_\textrm{BKT} \approx 0.148t$ is attainable at half filling when $U \approx 4.04t$ 
with $\mu = 0$ and $\Delta \approx 1.37t$. However, we also find that higher critical 
temperatures $T_\textrm{BKT}/t \approx \{0.169, 0.199, 0.231\}$ may be achieved, 
respectively, with $t'/t = \{ -0.1, -0.2, -0.3 \}$ at fillings $F \approx \{1.34, 1.40, 1.43\}$ 
when $U/t \approx \{3.40, 3.76, 4.19\}$. This is in agreement with our 
expectation that
$
T_\textrm{BKT} = \pi F (2-F) (t^2 + 6t'^2)/(2\sqrt{3} U)
$
increases with $t' \ne 0$ independently of its sign in the $U \gg t$ limit. Indeed, 
our numerical results benchmark very well with this analytic expression in the 
regime of its validity. Note that our result is considerably higher than the maximal 
value $T_\textrm{BKT} \approx 0.1t$ reported in the literature for 
$t' = -0.15t$~\cite{zhao06}. 

The discrepancy between our findings and the literature may well be caused 
by the use of a non-fully-self-consistent approach that is based on the numerical 
extraction of $D_0$ at $T = 0$~\cite{zhao06}. In addition, suspecting that the 
novel geometric contribution to the SF stiffness may partly be responsible 
for the apparent disagreement, we also present the relative weight 
$D_0^\textrm{geom}/D_0$ of this contribution for the same parameters. 
For instance, for $t'/t = \{0, -0.1, -0.2\}$, $D_0^\textrm{geom}/D_0$ 
reaches its maximum value $\{0.53, 0.54, 0.56\}$ at $F \approx \{1.0, 0.97, 0.89\}$ 
when $U/t \approx \{2.77, 2.76, 2.75\}$, and it is $\{0.50, 0.22, 0.15\}$ 
at the location of the maximal $T_\textrm{BKT}/t$. Thus, the geometric 
contribution is a non-monotonous function of $U$, and it accounts for a sizeable 
fraction of the SF stiffness in general reaching beyond $\%50$. 
In particular, Fig.~\ref{fig:F} shows that its role becomes more and more 
(less and less) critical at lower (higher) fillings with decreasing $t'/t$. 

In the $U \gg t$ limit, we may relate
$
D_0 = 4F (2-F) (t^2 + 6t'^2)/(\sqrt{3} U)
$
to the density $\rho_p$ and effective mass $m_p$ of the SF pairs through the
identity
$
D_0 = 4\hbar^2 \rho_p / m_p,
$
where 
$
\rho_p = 4F_c/(3\sqrt{3}a^2)
$
with 
$
F_c =  [\Delta^2/(4M)] \sum_{s \mathbf{k}} 
\mathcal{X}_{s \mathbf{k}}^2 / E_{s\mathbf{k}}^2
$
the filling of condensed pairs~\cite{iskin17}. This leads to $F_0 = F(2-F)/4$ 
as the filling of SF pairs whose effective mass $m_p = \hbar^2 U/[3(t^2 + 6t'^2)a^2]$ 
increases with $U$ but decreases with $t'$ in the $U \gg t$ limit. 
For completeness, the fraction $F_c/F_0$ of condensed particles is also shown 
in Figs.~\ref{fig:mu} and~\ref{fig:F} for the parameter regimes of interest. 
In comparison to the half-filling $F = 1$ case where half of the pairs or holes may 
at most be condensed with $F_0 \to 1/2$ in the $U \gg t$ limit, all of the particle 
(hole) pairs are condensed with $F_0 \to F/2$ ($F_0 \to 1-F/2$) in the low particle 
(hole) filling $F \to 0$ ($F \to 2$) limit.

\section{Discussion}
\label{sec:disc}

\begin{figure}[htbp]
\includegraphics[scale=0.38]{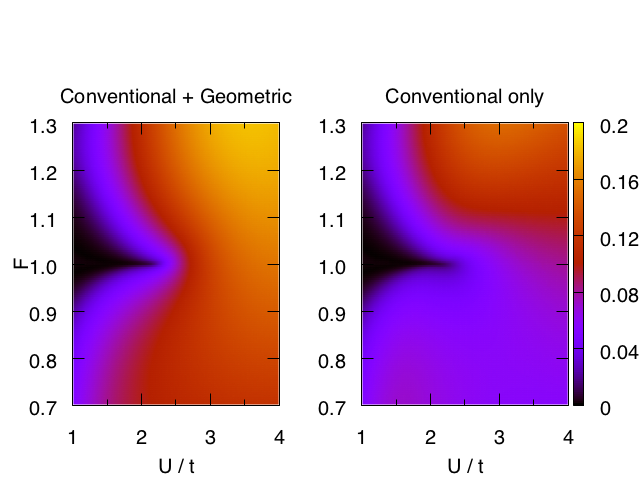}
\caption{(color online)
\label{fig:0_15}
The critical SF transition temperature $T_\textrm{BKT}/t$ is shown for the parameter 
regime of Ref.~\cite{zhao06} for $t' = -0.15 t$.
}
\end{figure}

In order to provide a better contextualization of our results, here we first calculate 
$T_\textrm{BKT}$ in a fully-self-consistent manner both with and without the 
geometric contribution $D_0^\textrm{geom}$, and benchmark these results directly 
with those of the literature. For this purpose, we set $t' = -0.15 t$ in Fig.~\ref{fig:0_15}, 
and present the resultant phase diagrams for precisely the same parameter 
window as the one that is shown in Ref.~\cite{zhao06}.

This figure clearly illustrates that exclusion of the $D_0^\textrm{geom}$ contribution 
from the universal BKT relation leads to a substantial reduction of $T_\textrm{BKT}$ 
for $U \gtrsim 2t$.This is in accordance with the discussion given above in 
Sec.~\ref{sec:bkt}, where $T_\textrm{BKT}$ is determined by the vanishing 
($\Delta \to 0$) of the BCS mean-field for $U \lesssim t$. In addition, since the latter 
phase diagram turns out to be quantitatively similar to that of Ref.~\cite{zhao06}, 
we speculate that the geometric contribution may not be taken fully into account by 
their approach. This is because while the self-consistent equations that are solved 
numerically for the pairing order parameter and particle filling are exactly the same 
in both works, there is one important difference in the way the SF stiffness is calculated.
That is our analytic expression for the SF stiffness is derived under the linear response 
theory, but Ref.~\cite{zhao06} extracts it numerically from the dispersion of the Goldstone 
mode which in turn is derived by considering the Gaussian fluctuations of the order 
parameter on top of the BCS ground state. Given our detailed benchmark, we believe 
these two approaches are equivalent for the $U \lesssim 2t$ region but not away 
from it for the honeycomb lattice.  As both approaches are routinely used to evaluate 
the SF stiffness and the related critical SF transition temperature, further work is needed 
to assess which method is more reliable and accurate, and whether they could be 
reconciled to some extent. In addition, a more detailed comparison with the state 
of the art unbiased numerical method is also highly desirable.

\section{Conclusions}
\label{sec:conc}

In summary, here we calculated the critical SF transition temperature of the 
attractive Hubbard model on a two-dimensional honeycomb lattice via a theoretical 
approach that is based on a self-consistent solution of the BCS mean-field 
theory for the stationary Cooper pairs and the universal BKT relation for the phase 
fluctuations. For instance, we found that the highest attainable $T_\textrm{BKT}$ 
is around $0.15t$ for the nearest-neighbor-hopping model, and that it increases 
quite rapidly with the inclusion of next-nearest-neighbor hoppings. In addition 
to the construction of the phase diagrams for a large window of model parameters, 
we also uncovered the critical role played by the quantum geometry of the 
underlying band structure. In particular, we found that the relative weight of the
quantum metric contribution to the SF phase stiffness is a non-monotonous 
function of the interaction strength, and that it is generally far from being 
negligible reaching beyond $\%50$. These findings arguably suggest that a 
SF Fermi gas that is loaded on a honeycomb 
lattice~\cite{tarruell12, uehlinger12, polini13, jotzu14, flaschner16} is one of the
ideal platforms for studying quantum geometric effects with cold atoms. 
The possible outcomes of such a realization would clearly have a broader 
impact in solid-state, condensed-matter and some other physics communities, 
given the modern surge of interest in the quantum topological and/or quantum 
geometrical concepts in general.

This line of work offers many extensions for future research. For instance, since 
the SF stiffness is determined by the ratio of the density and the mass of the SF 
carriers, we expect sizeable geometric contributions for those observables 
(e.g., sound velocity) that have explicit dependence on the mass of the two-body 
bound states or of the Cooper pairs on general grounds~\cite{iskin18b}. 
Given our findings for the geometric SF stiffness, we expect these observables 
to have a non-monotonous dependence on the interaction strength as well. 
This distinguishes the honeycomb lattice from the square-like Bravais lattices, 
for which the corresponding ground-state observables are known to evolve 
monotonously in the usual BCS-BEC crossover problem~\cite{nsr85, randeria92}. 
In addition to the honeycomb system, we are aware of other two-band systems 
with a non-trivial quantum geometry, exhibiting similar geometric effects. 
For instance, the Haldane-Hubbard model~\cite{torma17a}, Kane-Mele-Hubbard 
model~\cite{torma17a}, time-reversal-invariant Hofstadter-Hubbard 
model~\cite{torma15, iskin17}, and the spin-orbit coupled Fermi gases~\cite{iskin18a} 
are described by the single-particle and mean-field Hamiltonians that are of 
exactly the same form as the ones considered in this paper. Thus, there is no 
doubt that understanding the quantum metric effects on a honeycomb lattice 
may eventually have far reaching implications for a wide-class of two-band SFs.

\begin{acknowledgments}
The author acknowledges financial support from T{\"U}B{\.I}TAK.
\end{acknowledgments}

\end{document}